\documentclass[twocolumn]{bmcart}% 
\usepackage{amsthm,amsmath}
\usepackage{math-symbols} 
\usepackage[utf8]{inputenc} 
\usepackage{indentfirst}
\usepackage{booktabs}
\usepackage{color, xcolor}
\usepackage{soul}
\usepackage{mdframed}
\usepackage{lineno} 
\def\includegraphics{}

\startlocaldefs
\endlocaldefs

\begin{document}

%%% Start of article front matter
\begin{frontmatter}

\begin{fmbox}
\dochead{Research}


\title{Sound field reconstruction using neural processes with dynamic kernels}

%%%%%%%%%%%%%%%%%%%%%%%%%%%%%%%%%%%%%%%%%%%%%%

%%%%%%%%%%%%%%%%%%%%%%%%%%%%%%%%%%%%%%%%%%%%%%

\author[
  addressref={aff1},                
  email={zining.liang@mail.nwpu.edu.cn}   % email address
]{\fnm{Zining} \snm{Liang}}
\author[
  addressref={aff1},
  corref={aff1},
  email={wen.zhang@nwpu.edu.cn}
]{\fnm{Wen} \snm{Zhang}}
\author[
addressref = {aff2},
email ={thushara.abhayapala@anu.edu.au}
]{\fnm{Thushara D.}
\snm{Abhayapala}}

%%%%%%%%%%%%%%%%%%%%%%%%%%%%%%%%%%%%%%%%%%%%%%
%%                                          %%
%% Enter the authors' addresses here        %%
%%                                          %%
%% Repeat \address commands as much as      %%
%% required.                                %%
%%                                          %%
%%%%%%%%%%%%%%%%%%%%%%%%%%%%%%%%%%%%%%%%%%%%%%

\address[id=aff1]{%                           % unique id
  \orgdiv{Center of Intelligent Acoustics and Immersive Communications, School of Marine Science and Technology},             % department, if any
  \orgname{Northwestern Polytechnical University},          % university, etc
  \city{Xi'an},                              % city
  \cny{China}  
% country
}

\address[id=aff2]{%                           % unique id
  \orgdiv{Audio and Acoustic Signal Processing Group, College of Engineering and Computer Science},             % department, if any
  \orgname{The Australian National University},          % university, etc
  \city{Canberra},                              % city
  \cny{Australia}                                    % country
}
%%%%%%%%%%%%%%%%%%%%%%%%%%%%%%%%%%%%

\begin{abstractbox}

\begin{abstract} 
Accurately representing the sound field with high spatial resolution is critical for immersive and interactive sound field reproduction technology. To minimize experimental effort, data-driven methods have been proposed to estimate sound fields from a small number of discrete observations. In particular, kernel-based methods using Gaussian Processes (GPs) with a covariance function to model spatial correlations have been used for sound field reconstruction. However, these methods have limitations due to the fixed kernels having limited expressiveness, requiring manual identification of optimal kernels for different sound fields. In this work, we propose a new approach that parameterizes GPs using a deep neural network based on Neural Processes (NPs) to reconstruct the magnitude of the sound field. This method has the advantage of dynamically learning kernels from simulated data using an attention mechanism, allowing for greater flexibility and adaptability to the acoustic properties of the sound field. Numerical experiments demonstrate that our proposed approach outperforms current methods in reconstructing accuracy, providing a promising alternative for sound field reconstruction.
\end{abstract}


\end{abstractbox}
\end{fmbox}% uncomment this for two column layout

\end{frontmatter}

%%%%%%%%%%%%%%%%
%% Background %%
%%
\section*{1  Introduction}

Accurately describing the characteristics of a sound field, including its spatial, temporal, and spectral properties, is crucial for various applications, particularly spatial audio, which aims to create realistic auditory environments through loudspeakers or headphones~\cite{plinge2018six,cobos2022overview}. With recent advances in immersive and interactive sound field reproduction technologies, the ability to render dynamically variable sound fields that allow for listener and source movement within the audio scene has become increasingly important. While obtaining continuous spatial coverage measurements of a sound field over a large area is extremely challenging~\cite{witew2017sampling,koyama2021meshrir,kristoffersen2021deep,samarasinghe20123d}, sound field reconstruction offers a resourceful approach to estimate the sound field from a limited set of discrete observations. Such methods can help overcome the limitations of direct measurement techniques and enable realistic, immersive audio experiences in real-world applications.

General solutions for sound field reconstruction typically rely on conventional linear regression, where the sound field is measured at multiple points and represented as a linear combination of basis functions such as plane waves, cylindrical or spherical harmonics~\cite{ward2001reproduction,ueno2017sound,betlehem2005theory,7230263}. However, a large number of basis functions are needed to accurately represent sound fields over a large spatial region using conventional linear regression. 
Under specific acoustic assumptions, it is possible to represent the sound field using sparse representations, including plane-wave~\cite{verburg2018reconstruction} or spherical wave~\cite{9746391} expansions, and modal decomposition~\cite{das2021room,mignot2013low}, as well as equivalent source methods~\cite{lee2017use,tsunokuni2021spatial,antonello2017room}. Many of these techniques employ the principle of compressed sensing principles~\cite{donoho2006compressed} to estimate undersampled data for sound field reconstruction. 

In the field of infinite-dimensional analysis of sound fields using kernel ridge regression, different approaches have been proposed to address the issue of basis function truncation and kernel parameter optimization~\cite{8116652, 9632731, ribeiro2023kernel}. While some methods, such as the hierarchical kernel used in~\cite{caviedes2021gaussian}, require manual adjustment of the kernel parameters to align with the specific characteristics of the sound field, other works~\cite{9632731,ribeiro2023kernel} have focused on adaptive kernels, i.e. the usage of pre-defined kernels or sub-kernels with automatically adjusted parameters. It is important to note that these traditional methods rely on pre-defined kernel functions.

Recently, there are several data-driven methods utilizing neural networks (NN) for specific tasks within the field of sound field reconstruction~\cite{lluis2020sound, pezzoli2022deep, fernandez2023generative, shigemi2022physics, figueroa2023reconstruction, hahmann2021spatial}. Many of these methods are inspired primarily by image restoration and segmentation techniques in computer vision. For example, convolutional neural network (CNN) architectures including U-Net~\cite{lluis2020sound} was proposed for reconstructing room transfer functions~\cite{lluis2020sound}, physics-informed CNNs was proposed for reconstructing sound fields generated by point sources~\cite{shigemi2022physics}, and MultiResUNet was used for microphone array based room impulse response interpolation~\cite{pezzoli2022deep}. However, a significant limitation of these data-driven approaches is their potential to exhibit poor generalization to new acoustic scenarios. The reliance on large amounts of training data that fully cover the range of possible acoustic scenarios is a challenging requirement for data-driven methods.

Our research focuses on reconstructing various types of sound fields in the frequency domain across an entire spatial region, using a limited number of discrete observations. The work is based on Gaussian processes (GPs), which are powerful probabilistic models that
can be used to capture the spatial correlation in the
field by employing a kernel function and also to handle
the uncertainty associated with the field’s variations.
The work in~\cite{caviedes2021gaussian} presents a pioneering approach to using GPs for sound field reconstruction, demonstrating the significant potential of this technique. However, one crucial aspect that strongly influences the performance of GP models is the choice of kernel function. At the moment, there are still several unresolved questions regarding kernel selection. Firstly, pre-defined kernels may have limited expressiveness, requiring a comprehensive analysis of sound field properties and extensive experimentation to identify the most appropriate kernel and determine its parameters. Secondly, the current work has primarily focused on sound field reconstruction of far-field sources or sparsely distributed sources in reverberant rooms. The kernel functions used in prior work do not adequately capture near-field acoustic properties, resulting in poor sound field reconstructions in the near-field. Hence, there is potential for further exploration into various types of sound fields, such as near-field sources and standing waves, etc. 

In summary, identifying an appropriate kernel for various types of sound fields can be challenging, and deriving a kernel from scratch is often intractable. To address these issues, this paper proposes a novel data-driven approach to sound field reconstruction using Neural Processes (NPs)~\cite{garnelo2018neural}. NPs enable us to parameterize GPs using a deep neural network. In addition, we introduce dynamic kernels that can effectively adapt to the properties of diverse sound fields by leveraging attention mechanisms.

In this paper, the primary objective is to achieve an accurate reconstruction of sound field magnitudes using minimal observations that are arbitrarily and irregularly distributed. The paper is structured as follows. Section 2 provides a review of the GPs model, including commonly used kernel functions, and highlights the limitations of this model. Building on this, Section 3 presents the conceptual framework and neural network architecture details of the proposed approach using NPs. Section 4 outlines the training procedure and presents results on the reconstruction accuracy of the proposed method, in comparison with the conventional linear regression models and data-driven models.

\section*{2  Overview of GPs}

\subsection*{2.1 GPs methodology}

The problem is defined as reconstructing a sound field within a specific area of interest, using only a limited and finite set of observations, which are denoted as 
$\tilde{\mathbf{u}}=\left[\tilde{u}\left(\mathbf{r}_{1},\omega\right ), \ldots\tilde{u}\left(\mathbf{r}_{N},\omega\right)\right]$, where $\mathbf{r} \in \Omega$ is the spatial locations and $\omega$ is the angular frequency. Hereafter, $\omega$ is omitted for notation simplicity. The observed pressure $\tilde{u}(\mathbf{r})$ at a location $\mathbf{r}$ is represented as
%%%%%%%%
\begin{equation}\label{1}
\tilde{u}(\mathbf{r})=f(\mathbf{r})+e(\mathbf{r}),
\end{equation}
where the true sound field $f(\mathbf{r})$ cannot be directly observed or measured, and $e(\mathbf{r})$ denotes the measurement noise~\cite{caviedes2021gaussian}.

Assuming the sound field  in the space is a zero mean complex GP, that is the distribution of sound pressure within that space follows a complex Gaussian distribution
\begin{equation}\label{2}
	\tilde{u}(\mathbf{r}) \sim \mathcal{C G \mathcal { P }}\left(0, \kappa\left(\mathbf{r}, \mathbf{r}^{\prime}\right)\right),
\end{equation}
where the covariance function, or the kernel, $\kappa\left(\mathbf{r}, \mathbf{r}^{\prime}\right)$ of the sound pressures between the spatial locations of $\mathbf{r}$ and $\mathbf{r}^{\prime}$ is written as
\begin{equation}\label{3}
	\kappa\left(\mathbf{r}, \mathbf{r}^{\prime}\right)=\mathbb{E}\left[u(\mathbf{r}) u\left(\mathbf{r}^{\prime}\right)\right].
\end{equation}

The measurement noise in (\ref{1}) is also assumed complex Gaussian with zero mean
\begin{equation}\label{4}
	{e}(\mathbf{r}) \sim \mathcal{C G \mathcal { P }}\left(0, \kappa_{e}\left(\mathbf{r}, \mathbf{r}^{\prime}\right)\right).
\end{equation}

To predict the sound pressure at a new location $\mathbf{r}_{*}$, we need to compute the posterior distribution of $u_{*}(\mathbf{r})$ given the observed data $\tilde{u}(\mathbf{r})$ and the kernel parameters.
This can be done using the conditional distribution of a multivariate normal distribution~\cite{rasmussen2006gaussian}, 
\begin{equation}\label{5}
	 u_{*}(\mathbf{r})\mid \mathbf{r_{*}},\mathbf{r}, \tilde{\mathbf{u}}\sim \mathcal{C G \mathcal { P }}\left(\mu_{u_{*}\mid \tilde{\mathbf{u}}}(\mathbf{r}), \kappa_{u_{*}\mid \tilde{\mathbf{u}}}\left(\mathbf{r}, \mathbf{r_{*}}\right)\right),
\end{equation}
where $\mu_{u_{*}\mid \tilde{\mathbf{u}}}(\mathbf{r})$ is the predictive mean, and $\kappa_{u_{*}\mid \tilde{\mathbf{u}}}\left(\mathbf{r}, \mathbf{r_{*}}\right)$ is the kernel between the observed position and predictive position.

The optimal sound field reconstruction is the posterior mean in (\ref{5}), that is
\begin{equation}\label{6}
	 \mu_{u_{*} \mid \tilde{\mathbf{u}}}(\mathbf{r})=\boldsymbol{\kappa}^{\mathrm{H}}(\mathbf{K}+\boldsymbol{\Sigma})^{-1} \tilde{\mathbf{u}},
\end{equation}
where the kernel, $\boldsymbol{\kappa}=\left[\kappa\left(\mathbf{r}_{1}, \mathbf{r}_{*}\right) \cdots \kappa\left(\mathbf{r}_{N},\mathbf{r}_{*}\right)\right]$, is the spatial correlation function between the $N$ observed pressures and the predictive locations $\mathbf{r_{*}}$, and the covariance
matrices $\mathbf{K}$ and $\boldsymbol{\Sigma}$ are defined as\cite{fernandez2021reconstruction}
\begin{equation}\label{7}
	 \begin{array}{l}
\boldsymbol{\Sigma}=\mathbb{E}\left[\mathbf{e e}^{\mathrm{H}}\right], \\
\mathbf{K}=\mathbb{E}\left[\mathbf{f} \mathbf{f}^{\mathrm{H}}\right].
\end{array}
\end{equation}

Obviously, the kernel function, which models the spatial correlation between the sound pressure measurements, is a crucial part of sound field reconstruction using GP. The choice of kernel function can have a significant impact on the accuracy and efficiency of the sound field reconstruction.

\subsection*{2.2 Kernel functions}

Kernels for sound field representation are typically categorized based on their properties of stationarity and isotropy. It is vital to choose or develop a kernel function that aligns with the characteristics of the sound field in GP methodology. For instance, a diffuse field demonstrates stationary and isotropic spatial correlation, while a plane wave field presents stationary but anisotropic spatial correlation. Below are some frequently applied kernel functions in audio and acoustics research~\cite{2011Gaussian}.

\subsubsection*{2.2.1 RBF kernels}

The definition of the isotropic radial basis function (RBF) kernel is
\begin{equation}\label{8}
	 	\kappa_{RBF_{i}}\left(\mathbf{r}, \mathbf{r}^{\prime}\right)=\alpha^{2} \exp \left(-\frac{1}{2 \rho^{2}}\|\boldsymbol{\delta}\|^{2}\right),
\end{equation}
where $\alpha$ is the scaling factor that adjusts the kernel functions to match the size of the data, $\rho$ is the length scale defining the decay rate of the kernel, and $\boldsymbol{\delta} \triangleq \mathbf{r}-\mathbf{r}^{\prime}$ is the euclidean distance between two points.

The definition of the anisotropic RBF kernel is
\begin{equation}\label{9}
	\kappa_{RBF_{a}}\left(\mathbf{r}, \mathbf{r}^{\prime}\right)=\alpha^{2} \exp \left(-\frac{1}{2} \sum_{l=1}^{L} \frac{\left\|\mathbf{u}_{l}^{\mathrm{T}} \boldsymbol{\delta}\right\|^{2}}{\rho_{l}^{2}}\right),
\end{equation}
where the unitary vector $\mathbf{u}_{l} \in \mathbb{R}^{D}$ defines  the $l$th direction and $\rho_{l}$ is the length scale of the corresponding direction.

The definition of the periodic RBF kernel, which is derived from Eq.(\ref{9}), gives
\begin{equation}\label{10}
	\kappa_{RBF_{p}}\left(\mathbf{r}, \mathbf{r}^{\prime}\right)=\alpha^{2} \exp \left(-\sum_{l=1}^{L} \frac{1}{2 \rho_{l}^{2}} \sin ^{2}\left(\frac{k\left\|\mathbf{u}_{l}^{\mathrm{T}} \boldsymbol{\delta}\right\|}{2}\right)\right),
\end{equation}
where the kernel repeats every wavelength $\lambda=2 \pi / k$.

\subsubsection*{2.2.2 The plane waves kernels}

Plane-wave expansions serve as a widely used method in sound field reconstruction. By decomposing the sound field into a sum of plane waves with varying amplitudes, directions, and frequencies, it becomes possible to reconstruct the field by determining their respective amplitudes and phases~\cite{schmid2021spatial,antonello2017room,mignot2013low,jacobsen2013fundamentals}. That is, at the wavenumber $k$, the field at any point in space $r$ can be expressed as
\begin{equation}\label{11}
	f(\mathbf{x})=\sum_{l=1}^{L} w_{l} \mathrm{e}^{-\mathrm{j} \mathbf{k}_{l}^{\mathrm{T}} \mathbf{r}},
\end{equation}
where $w_{l}$ are unknown weights, $\mathrm{e}^{-\mathrm{j} \mathbf{k}_{l}^{\mathrm{T}} \mathbf{r}}$is the elementary wave function, and $\mathbf{k}_{l}=k\mathbf{u}_{l}$ is the wavenumber vector.

If the weights $w_{l}$ are also modeled as a complex Gaussian process such that
\begin{equation}\label{12}
w_{l} \sim \mathcal{C G \mathcal { P }}\left(0, \sigma_{l}^{2}\right),
\end{equation}
the kernel for the sound field that is generated by multiple sound sources~\cite{nolan2018wavenumber,fernandez2016sound} is defined as
\begin{equation}\label{13}
\kappa_{m}\left(\mathbf{r}, \mathbf{r}^{\prime}\right)=\sigma_{\mathbf{w}}^{2} \sum_{l=1}^{L} \mathrm{e}^{-\mathrm{j}\mathbf{k}_{l}^{\mathrm{T}} \boldsymbol{\delta}},
\end{equation}
where the weights $w_{l}$ share a same variance $\sigma_{\mathbf{w}}^{2}$.
For a special case that
the sound field is generated by only a few sources, which is normally characterized as sparse~\cite{gemba2017multi,8649645}, and the kernel is defined as
\begin{equation}\label{14}
\kappa_{s}\left(\mathbf{r}, \mathbf{r}^{\prime}\right)=\sum_{l=1}^{L} \sigma_{l}^{2} \mathrm{e}^{-\mathrm{j} \mathbf{k}_{l}^{\mathrm{T}} \boldsymbol{\delta}},
\end{equation}
where the variances of the weights $w_{l}$ are independent, and $\sigma_{l}$ are considered as inverse gamma distributed~\cite{murphy2012machine}
\begin{equation}\label{15}
\sigma_{l} \sim \Gamma^{-1}(a, b)=\frac{b^{a}}{\Gamma(a)}\left(1 / \sigma_{l}\right)^{a+1} \exp \left(-b / \sigma_{l}\right),
\end{equation}
where $a > 0$ is the shape parameter and $b > 0$ is the scale parameter of the density function. With a fixed prior $a$, smaller values of $b$ promote sparser solutions.

The concept of the hierarchical kernel $\kappa_{h}$ is introduced in \cite{caviedes2021gaussian}. In order to adapt to both normal and sparse sound field, the parameters $\sigma_{l}$ in (\ref{15}) is defined as
\begin{equation}\label{16}
\sigma_{h} \sim \Gamma^{-1}(1, b), \quad b \sim \mathcal{N}\left(\mu_{b}, \sigma_{b}\right).
\end{equation}

\subsubsection*{2.2.3 The diffuse field kernel}

For a diffuse field driven by a pure tone, the spatial correlation and coherence can be modeled by the superposition of an infinite number of random phase plane waves~\cite{jacobsen2013fundamentals}. That is, the diffuse field kernel function corresponding to (\ref{11}) in the limit $L \rightarrow \infty$ is written as follows,
\begin{equation}\label{17}
	\kappa_{f}\left(\mathbf{r}, \mathbf{r}^{\prime}\right)=\sigma_{\mathbf{w}}^{2} \lim _{L \rightarrow \infty} \sum_{l=1}^{L} \mathrm{e}^{-\mathrm{j} \mathbf{k}_{l}^{\mathrm{T}} \boldsymbol{\delta}}.
\end{equation}

For the two-dimensional case, the kernel in (\ref{17}) is the zeroth-order Bessel function
\begin{equation}\label{18}
	\kappa_{b}\left(\mathbf{r}, \mathbf{r}^{\prime}\right)=\frac{\sigma_{\mathbf{w}}^{2}}{2 \pi} \int_{-\pi}^{\pi} \mathrm{e}^{-\mathrm{j} k\|\boldsymbol{\delta}\| \cos \varphi} \mathrm{d} \varphi=\sigma_{\mathbf{w}}^{2} \mathrm{~J}_{0}(k\|\boldsymbol{\delta}\|).
\end{equation}

In summary, when attempting sound field reconstruction using GPs, it is necessary to understand the characteristics of the sound field and select the appropriate kernel function. Once the optimal kernel function is determined, (\ref{6}) can be utilized to obtain the predictive sound field pressure.
However, if there is no suitable kernel function available, a custom kernel may need to be derived. Nevertheless, developing a kernel that can effectively adapt to diverse acoustic environments can be challenging, particularly when dealing with complex sound fields. Additionally, estimating the optimal hyperparameters of the kernel through numerous experiments can be a time-consuming process.

\section*{3  Proposed method}

In this work, we propose a novel approach to automatically obtain the optimal kernel from the magnitude of the sound field data for reconstruction, using a data-driven model based on NPs with attention mechanisms. Our proposed model generates dynamic kernels that can adapt to the unique properties of various sound fields and defines distributions over sound field functions similar to GPs. This combination provides a probabilistic, data-efficient, flexible, and computationally efficient solution for optimal kernel selection.

In this section, we first detail the overall architecture of our method in Section 3.1, and then introduce the proposed two-stream encoder and the efficient and lightweight decoder in Section 3.2 and Section 3.3, respectively.

\subsection*{3.1 Architecture}

\begin{figure*}[ht!]
  \centering
  \includegraphics[scale=0.5]{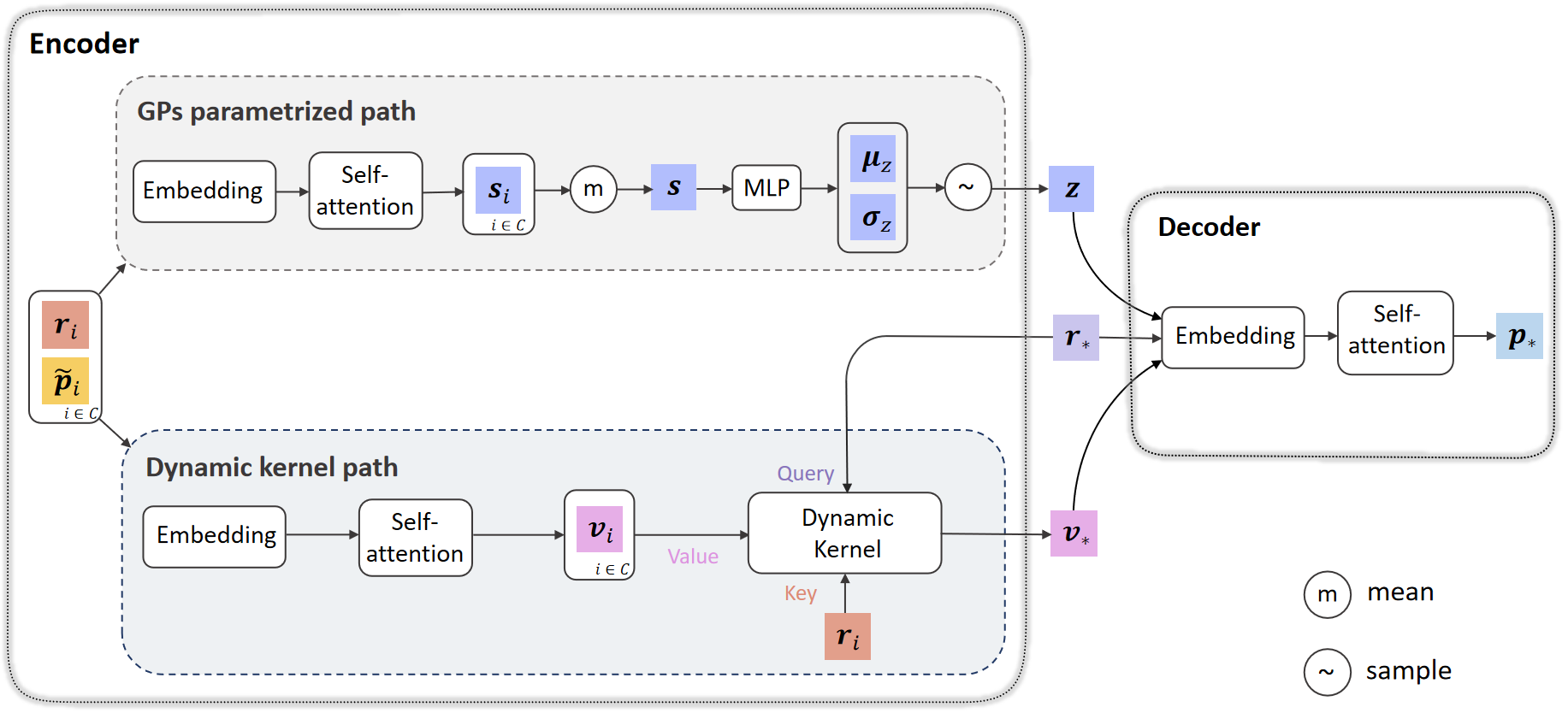}
  \caption{Schematic diagram of the neural network architecture proposed for sound field reconstruction.}
 \label{fig1}
\end{figure*}

As shown in Figure \ref{fig1}, the proposed model is composed of an encoder and a decoder. Specifically, the encoder contains two paths: a GPs parameterized path, which models the global structure of the stochastic process realization, and a dynamic kernel path, which captures the spatial correlation between observations and predictions.

The encoder takes a limited set of observed sound field magnitude measurements along with their corresponding locations ${(\mathbf{r},\tilde{\mathbf{p}})}_{i\in C(0, N)}$ as input, where $\mathbf{p}=\left| \mathbf{u} \right|$ and $C$ denotes the set of integers from $0$ to $N$. Within the GPs parameterized path, the encoder outputs a latent variable $\mathbf{z}$, which encodes the global structure and uncertainty of the sound field distributions in the function space. 
In the dynamic kernel path, given the target location $\mathbf{r_{*}}$, the dynamic kernel mechanism outputs a correlation-specific representation $\mathbf{v}_{*}$. Since the dynamic kernel models the spatial correlation between observations and predictions using differentiable attention, which cannot be analytically obtained and acts as an implicit kernel, we visualize it in section 4.6.

The decoder takes the latent variable $\mathbf{z}$, the correlation-specific representation $\mathbf{v}_{*}$, and the target location $\mathbf{r_{*}}$ as input, and produces the predictive sound field magnitude $\mathbf{p}_{*}$ of the target location. This process can be understood as analogous to reconstructing the sound field using an appropriate kernel, utilizing the neural network to carry out the calculation described by (\ref{5}).

\subsection*{3.2 Encoder}
In this section, we introduce the structure and mechanism of the encoder with the two distinct paths.

\subsubsection*{3.2.1 GPs parameterized using NPs}

The GPs parameterized path is designed to learn distributions over sound field functions from observations. 
To represent a GP using a neural network, we assume that $F(x) \sim \mathcal{G \mathcal { P }}\left(\mu, \sigma\right)$ can be parameterized by a high-dimensional random vector $\mathbf{z}$, i.e., the latent variable~\cite{garnelo2018neural}. We can then write $F(x) = g(x, \mathbf{z})$ for some fixed and learnable function $g$, where $\mathbf{z}$ models different realizations of the data-generating GPs~\cite{kim2019attentive}. The motivation for introducing $\mathbf{z}$ is to enable our model to capture different types of sound fields.

In the GPs parameterized path, the observed sound field magnitudes in the frequency-spatial domain ${(\mathbf{r},\tilde{\mathbf{p}})}_{i}$ are embedded from the input space to the representation space using fully-connected layers with Gaussian Error Linear Unit (GELU)~\cite{hendrycks2016gaussian} activation functions. In our approach, we incorporate a self-attention (SA) mechanism~\cite{vaswani2017attention}, denoted as $\mathbf{s}_{i}=SA(\mathbf{r}_{i},\tilde{\mathbf{p}}_{i})$, to model higher-order interactions within the sound field. The SA mechanism allows us to capture the interactions among the observations, enabling the learning of global structural features of the sound field, and obtaining richer representations of the observations. The mean aggregator is used to combine the features as $\mathbf{s}=m\left(\mathbf{s}_{i}\right)$, and generate a single global representation by 
Multi-Layer Perceptron (MLP), which parameterizes the latent distribution $\mathbf{z} \sim \mathcal{GP}(\boldsymbol{\mu}_{z}, \boldsymbol{\sigma}_{z})$. Finally, each sample of $\mathbf{z}$ corresponds to one realization of the GPs, capturing the global uncertainty.

In summary, the GPs parameterized path learns the mapping from the observed data to the latent distribution of the GPs, representing Eq.(\ref{5}) by the neural network. Following this framework, the kernel function is not explicitly defined but is learned through the neural network's parameters, which is described in detail below. 

\subsubsection*{3.2.2 Dynamic kernel based attention mechanism}

In GPs, the kernel function captures the relationship between pairs of inputs by computing the dot product between their corresponding feature maps. Here, the kernel is defined as $\kappa\left(x,x^{\prime}\right)=\left\langle\Phi(x), \Phi\left(x^{\prime}\right)\right\rangle=\Phi(x)^{\top}\Phi\left(x^{\prime}\right)$, where $\Phi$ represents the feature map that maps the inputs into a higher-dimensional feature space. The advantage of using such a kernel is that it allows us to design algorithms based on dot-product spaces~\cite{hofmann2008kernel}. In our approach, we introduce a dynamic kernel mechanism inspired by the Scaled Dot-Product Attention (SDPA)~\cite{vaswani2017attention}. This dynamic kernel mechanism enables us to model the spatial correlation presented in diverse sound fields. More specifically, the target location $\mathbf{r_{*}}$ is treated as a query, while the observations ${(\mathbf{r},\mathbf{v})}_{i}$ are treated as key-value pairs. Here, $\mathbf{v}_{i}$ represents the transformation of $\tilde{\mathbf{\mathbf{p}}}_{i}$ into a higher-dimensional space through embedding. Similarly, both $\mathbf{r}_{i}$ and $\mathbf{r_{*}}$ undergo embedding within the dynamic kernel mechanism. The SDPA mechanism allows us to calculate weights that determine the correlation of each observation with respect to the target location, enabling accurate prediction of the sound field magnitude $\mathbf{p}_{*}$ at the target location.%By incorporating the dynamic kernel mechanism based on SDPA, the dynamic kernel path is able to capture the spatial correlation present in diverse sound fields. This results in more robust and accurate predictions of the sound field magnitude at the target location.}

Suppose we have $n$ key-value pairs arranged as matrices $\mathbf{R} \in \mathbb{R}^{n \times d_{r}}$, $\mathbf{V} \in \mathbb{R}^{n \times d_{v}}$, and $m$ queries $\mathbf{R_*} \in \mathbb{R}^{m \times d_{r}}$. The dynamic kernel mechanism calculates correlation weights $\boldsymbol{\kappa}_{d}$ by taking
the dot-product of the queries and keys scaled by $d_{r}$, i.e., the kernel form~\cite{tsai2019transformer}, and assigns $\boldsymbol{\kappa}_{d}$ to $\mathbf{V}$ to obtain the output $\mathbf{V}_*$, which gives
\begin{equation}\label{19}
\begin{aligned}
\boldsymbol{\kappa}_{d} &= \operatorname{softmax}\left(\mathbf{R_*} \mathbf{R}^{\top} / \sqrt{d_{r}}\right), \\
	\mathbf{V_*}&=\boldsymbol{\kappa}_d \mathbf{V} \in \mathbb{R}^{ d_{v}}.
\end{aligned}
\end{equation}
	
% Multi Dynamic Kernal (MDK)
In addition to using a single dynamic kernel, we further propose using a multi-dynamic kernel to achieve linear smoother query values \cite{vaswani2017attention, tsai2019transformer}.As shown in Eq.(\ref{20}), the multi-dynamic kernel is obtained by the sum of $h$ kernels mapping with different weights $\mathbf{W}$, defined by
\begin{equation}\label{20}
\begin{aligned}
\boldsymbol{\kappa}_i &= \operatorname{softmax}\left(\mathbf{R_*} \mathbf{W}_{*i} (\mathbf{R}\mathbf{W_i})  ^{\top} / \sqrt{d_{k}}\right),  \\
\mathbf{V_*} & =\left(\boldsymbol{\kappa}_{1}, \ldots, \boldsymbol{\kappa}_{h}\right) \mathbf{V } \in \mathbb{R}^{d_{v}},i\in [1,h].
\end{aligned}
\end{equation}

For each target location $\mathbf{r_*}$, the dynamic kernel generates an attention map between $\mathbf{r_*}$ and observations ${(\mathbf{r},\tilde{p})}_{i}$, which are totally learned from the data. This allows our proposed model to make more accurate predictions in environments with different acoustic properties. The visualization of this part is shown in section 4.6.

\subsection*{3.3 Decoder}

The decoder takes the latent variable $\mathbf{z}$, the correlation-specific representation $\mathbf{v}_{*}$, and the target location $\mathbf{r}_{*}$ as input. We define a Gaussian likelihood to describe the decoder, that is
\begin{equation}\label{21}
\pi({\mathbf{p}}_{*} \mid \mathbf{z}, \mathbf{v}_{*}, \mathbf{r}_{*})=\mathcal{N}\left(\mathbf{p}_{*} \mid g_{\theta}(\mathbf{r}_{*},\mathbf{z}),\mathbf{v}_{*},  \tau^{-1} \mathbf{I}\right),
\end{equation}
where $\mathbf{z}$ is a global latent variable, $g_{\theta}(\mathbf{r}_{*},\mathbf{z})$ is a decoder function to generate a prediction for target sound field magnitude ${\mathbf{p}}_{*}$ at a location $\mathbf{r}_{*}$, which is implemented as a deep neural network with parameters $\theta$, and $\tau^{-1}$ is the variance of observation noise\cite{rudner2018connection,garnelo2018neural}. Specifically, the likelihood $\pi({\mathbf{p}}_{*} \mid \mathbf{z}, \mathbf{v}_{*}, \mathbf{r}_{*})$ is defined as a factorized Gaussian distribution  across the predictions $(\mathbf{r}_{*}, \mathbf{p}_{*})$ with mean and variance determined by $\mathbf{z}$ and correlation-specific representation $\mathbf{v}_{*}$.

To generate the predictive sound field magnitude $\mathbf{p}_{*}$, the proposed model is defined by
\begin{equation}\label{22}
\pi({\mathbf{p}}_{*},\mathbf{z} \mid  \mathbf{v}_{*}, \mathbf{r}_{*})=\pi(\mathbf{z}\mid\mathbf{s})\mathcal{N}\left(\mathbf{p}_{*} \mid g_{\theta}(\mathbf{r}_{*},\mathbf{z}),\mathbf{v}_{*}, \tau^{-1} \mathbf{I}\right).
\end{equation}

Since the conditional prior $\pi(\mathbf{z}\mid\mathbf{s})$ in Eq.(\ref{22}) is intractable, it is approximated using the variational posterior~\cite{garnelo2018neural}
\begin{equation}\label{23}
q(\mathbf{z} \mid \mathbf{s})=\mathcal{N}\left(\mathbf{z} \mid  \boldsymbol{\mu}_{z}\left(m\left(\mathbf{s}_{i}\right)\right), \boldsymbol{\sigma}_{z}\left(m\left({\mathbf{s}}_{i}\right)\right)\right).
\end{equation}
where $m(\cdot)$ is a mean aggregator function, and $\boldsymbol{\mu}_{\omega}(\cdot)$ and $\boldsymbol{\sigma}_{\omega}(\cdot)$ parameterize a normal distribution from which $\mathbf{z}$ is sampled.

\subsection*{3.4 Loss function}
The parameters of the encoder and decoder are learned by maximizing the evidence lower-bound (ELBO), 
\begin{equation}\label{24}
\begin{aligned}
L_{\mathrm{ELBO}} = & -\mathbb{E}_{q\left(\mathbf{z} \mid {s}_{*}\right)}\left[\log \pi\left(\mathbf{p}_{*} \mid \mathbf{z}, \mathbf{v}_{*},\mathbf{r}_{*}\right)\right] \\
& +\operatorname{KL}\left(q\left(\mathbf{z} \mid \mathbf{s}_{*}\right) \| q\left(\mathbf{z} \mid \mathbf{s}\right)\right).
\end{aligned}
\end{equation}
The objective function consists of two terms. The first term is the reconstruction error (RE), which is equivalent to the mean squared error (MSE)~\cite{rey2019diffusion}. We denote this term as $L_{D}$, and it measure the discrepancy between the predicted output  $\mathbf{p}_{*}$ and the corresponding ground truth $\mathbf{p}_{\bullet}$. The MSE is computed over all the elements, denoted as $\mathcal{N}$. The second term is called the Kullback–Leibler(KL) divergence\cite{kullback1951information}, which is a measure of dissimilarity between two probability distributions. It quantifies the difference between the distribution of observed data $q\left(\mathbf{z} \mid \mathbf{s}\right)$ and the distribution of unobserved data $q\left(\mathbf{z} \mid \mathbf{s}_{*}\right)$ during the training process. 
\begin{equation}\label{25}
L_{\mathrm{D}}=\frac{1}{\mathcal{N}} \sum_{i \in \mathcal{N}}\left|\mathbf{p}_{*}\left(\boldsymbol{r}_{i}\right)-\mathbf{p}_{\bullet}\left(\boldsymbol{r}_{i}\right)\right|^{2} .
\end{equation}
To achieve a balance between data reconstruction and meaningful representation learning, we assign equal weights to both terms during training.

\section*{4  Simulation experiments}

We evaluated the performance of our proposed sound field reconstruction model in comparison to the GP and data-driven models. The sound fields we reconstructed included both spatially stationary and non-stationary fields, such as a diffuse field and point sources in the near-field. Additionally, we reconstructed simulated room transfer functions (RTFs) using the image source method~\cite{habets2006room} and modal theory~\cite{jacobsen2013fundamentals}. Our reconstruction was carried out on a two-dimensional grid composed of 32 by 32 uniformly spaced points along the relevant dimensions. The absolute distance between input points is determined by the room size. Specifically, the distance between points along the x-axis is $l_x/32$, and the distance between points along the y-axis is $l_y/32$. To ensure scale independence in the learning process, it is common to standardize the input for each frequency. This standardization involves transforming the input values such that they have a mean of 0 and a standard deviation of 1.
%The absolute distance between points depends on the room size. For example, in a room with dimensions of $l_x \times  l_y$, input points will be at a distance of $l_x/32$ and $l_y/32$. For scale-independent learning, the input is standardized for each frequency to have a mean of 0 and a standard deviation of 1.}

\subsection*{4.1 Evaluation metrics}

We use two metrics to evaluate the performance of our models. The first metric is the normalized mean square error (NMSE) between the ground truth $\mathbf{p_{\bullet}}$ and the predictions $\mathbf{p_{*}}$ for each frequency point $k$, which is calculated as follows,
%%%%%%%%%
\begin{equation}\label{27}
\mathrm{NMSE}_{k}=\frac{1}{N} \sum_{i=1}^{N} \frac{\left\|{p}_{\bullet}\left(\mathbf{r}_{i},\omega_{k}\right )-{p_{*}}(\mathbf{r}_{i},\omega_{k}\right )\|_{2}^{2}}{\|{p}_{\bullet}\left(\mathbf{r}_{i},\omega_{k}\right )\|_{2}^{2}}.
\end{equation}

The second metric is the Modal Assurance Criterion (MAC)~\cite{pastor2012modal} for each frequency point $k$, which is defined as follows,
\begin{equation}\label{28}
\mathrm{MAC}_{k}=\frac{\left\|\mathbf{p}_{\bullet k}^{\mathrm{T}} \mathbf{p}_{* k}\right\|_{2}^{2}}{\left(\mathbf{p}_{\bullet k}^{\mathrm{T}} \mathbf{p}_{\bullet k}\right)\left(\mathbf{p}_{* k}^{\mathrm{T}} \mathbf{p}_{* k}\right)}.
\end{equation}
The MAC measure evaluates the level of spatial similarity by determining how well the model predicts the overall shape of the pressure distribution in the sound field for each frequency point. The MAC values range from 0 (indicating maximum dissimilarity) to 1 (representing identical shapes), providing a quantitative measure of the quality of the model's predictions. 

\subsection*{4.2  Training procedure}

Our proposed model can be trained end-to-end on simulated data. To optimize the model, we use the Adam optimizer \cite{kingma2014adam} and train it for 300 epochs. The base learning rate is initially set to 1e-4 and decays to 1e-5 after 200 epochs. Moreover, to achieve better performance and stability during the training process, we implement an exponential warm-up strategy throughout the first 20 epochs.

\subsection*{4.3 Spatially stationary field}

\begin{figure*}[ht!]
  \centering
  \includegraphics[scale=0.55]{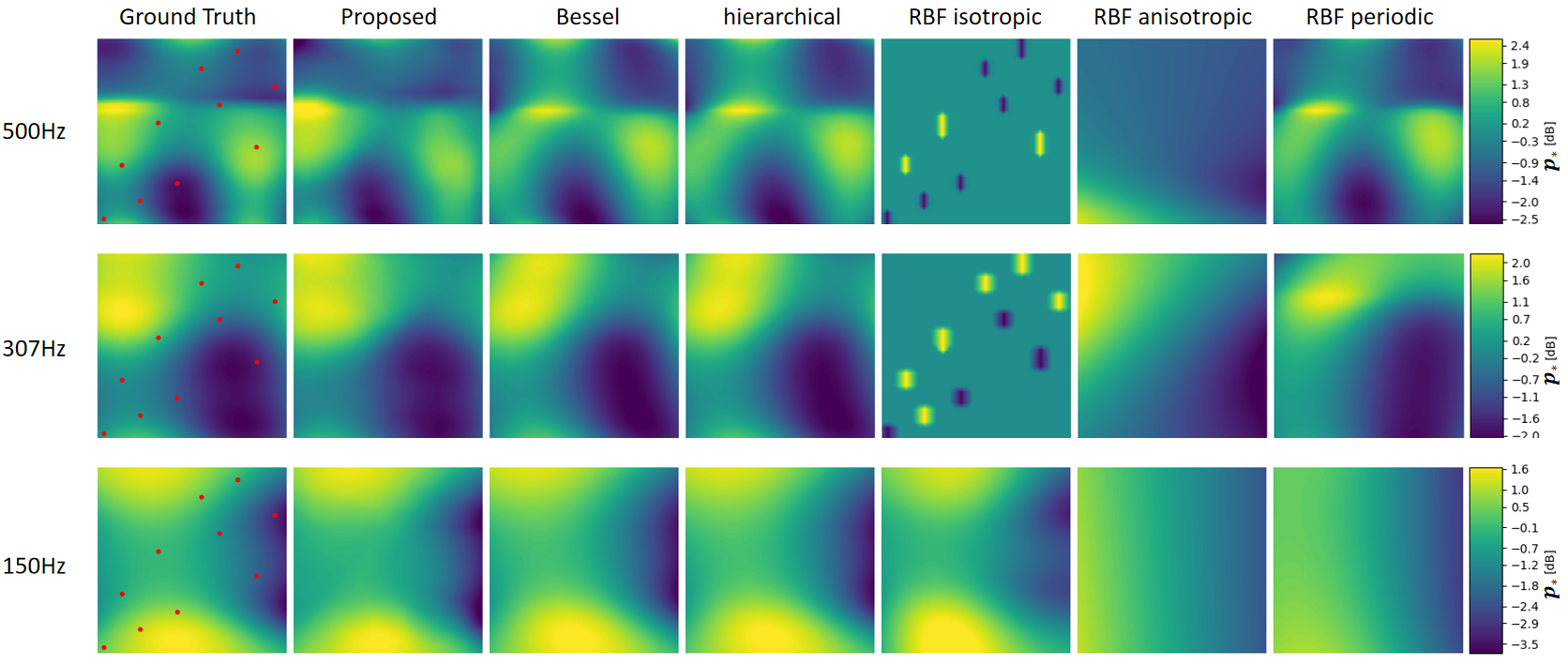}
  \caption{Reconstructed diffuse-field magnitudes of different frequencies given 10 observations arbitrarily placed. The red dots indicate the location used for reconstructing predicted sound field magnitude.} 
 \label{fig2}
\end{figure*}

\begin{table*}[ht]
\caption{The mean of NMSE and MAC of diffuse-field test dataset of different frequencies given 10 observations arbitrarily placed.}\label{tab1}
\begin{tabular*}{\textwidth}{@{\extracolsep\fill}lcccccc}
\toprule%
& \multicolumn{3}{@{}c@{}}{NMSE(dB)} & \multicolumn{3}{@{}c@{}}{MAC} \\\cmidrule{2-4}\cmidrule{5-7}%
Method & 150Hz & 307Hz & 500Hz & 150Hz & 307Hz & 500Hz \\
\midrule
\textbf{Proposed}  & \textbf{-22.8639} & \textbf{-19.6237} & \textbf{-11.3852} & \textbf{0.9948}& \textbf{0.9893} & \textbf{0.9278}\\
Bessel  & -9.5050 & -7.6729  & -1.2261  & 0.9227& 0.9642 & 0.7874\\
hierarchical  &-10.2242 & -7.9026  & -1.4814  & 0.9305 & 0.9648  &0.7877\\
RBF isotropic  & -7.1763 & -0.0536  & -0.0431  &0.9224& 0.0080 & 0.0112\\
RBF anisotropic  &2.9634 & 1.3958 & 2.4845  &0.8075 & 0.1106  & 0.0063\\
RBF periodic  & -2.0407 & 1.4845 & 1.8934 &0.8606 & 0.7885  & 0.6728\\
\bottomrule
\end{tabular*}
\end{table*}

In this section, we explore the reconstruction of the diffuse field, which is modeled by the superposition of an infinite number of random phase plane waves, as shown in Eq.(\ref{17}). This type of sound field is particularly relevant to the sound field present in reverberation rooms \cite{2019Experimental}.

To evaluate the performance of our proposed model, we conduct experiments on simulated data. Specifically, we estimate the sound field magnitudes in the frequency band [30, 500] Hz on a 32 by 32 grid, given 10 observations arbitrarily placed. The simulated data is generated by using $m$ plane waves with unit magnitude and random phase, i.e., $\angle \mathbf{u}_{l} \sim \mathcal{U}{[0,2 \pi)}$ and random direction of propagation, i.e., $ \mathbf{k}_{l} \sim \mathcal{U}{[-k,k]}$. Here, $m$ is randomly sampled from the range of $m \in (1000, 3000)$. To train our proposed model, we use a diverse set of 8000 diffuse fields according to the above parameter settings.

In order to evaluate the effectiveness of our proposed model, we compare it against GPs with different kernels, including the Bessel kernel, hierarchical kernel, and RBF kernels. The prior densities of parameters in Eq. (\ref{8}) - (\ref{10}) are defined as $\alpha \sim \mathcal{N}(0,1)$, $\rho$ and $\rho_l \sim \Gamma^{-1}\left(a_{\rho}, b_{\rho}\right)$, where $a_{\rho}=5$ and $b_{\rho}=5$. For the hierarchical kernel in Eq.(\ref{16}), the parameters are set as $b=10^{-b_{\log}}$ and $b_{\log} \sim \mathcal{N}(2,1)$ \cite{caviedes2021gaussian}. In order for the mean magnitude of the fields to be 1 Pa, the fields are normalized. The parameters settings and scaling aligh with the original work~\cite{c2021}.

In Table \ref{tab1}, we present the mean performance of our proposed model compared to GPs on a diverse set of 1000 diffuse fields. The results clearly demonstrate that our model exhibits significantly improved reconstruction performance. To provide a detailed visualization of the reconstruction process, we selected a sound field from the test set. Figure \ref{fig2} depicts the sound field magnitudes of the reconstructed data at various frequencies. The Bessel kernel performs relatively well due to its aptitude to coincide with the diffuse field. The hierarchical kernel exhibits a certain level of adaptability to the property of the sound field, enabling it to capture the structure of the diffuse field. However, in regions where there are no observations, such as the upper left corner, all kernels poorly extrapolate the sound field, particularly at 500Hz. This phenomenon highlights the limitations of the GPs method in accurately capturing the complex behavior of the sound field in sparsely sampled regions.

In comparison, the proposed model achieves the best performance due to the proposed attention-based dynamic kernel mechanism, which enables the model to effectively capture the global sound field and obtain richer representations. This enhances the model's overall performance, enabling it to outperform other approaches.

\subsection*{4.4 Spatially non-stationary field}
In this section, we discuss the process of reconstructing the sound field in the near-field created by multiple point sources. This type of sound field is particularly relevant to the direct component of the Room Impulse Response (RIR)~\cite{allen1979image, ribeiro2023kernel}. The direct component of the RIR provides critical information about the room geometry~\cite{dokmanic2013acoustic}.

To train our model, we created a dataset consisting of 8000 simulated sound fields. Each field is composed of a random number of point sources, denoted as $j \sim \mathcal{U}[1,6]$, which are randomly distributed. Each point source is positioned at a radial distance, represented by $d \sim \mathcal{U}(\lambda, 3\lambda)$, from the central point of the reconstruction area. The parameters of GPs method are set as section 4.3.

\begin{figure*}[ht!]
  \centering
  \includegraphics[scale=0.55]{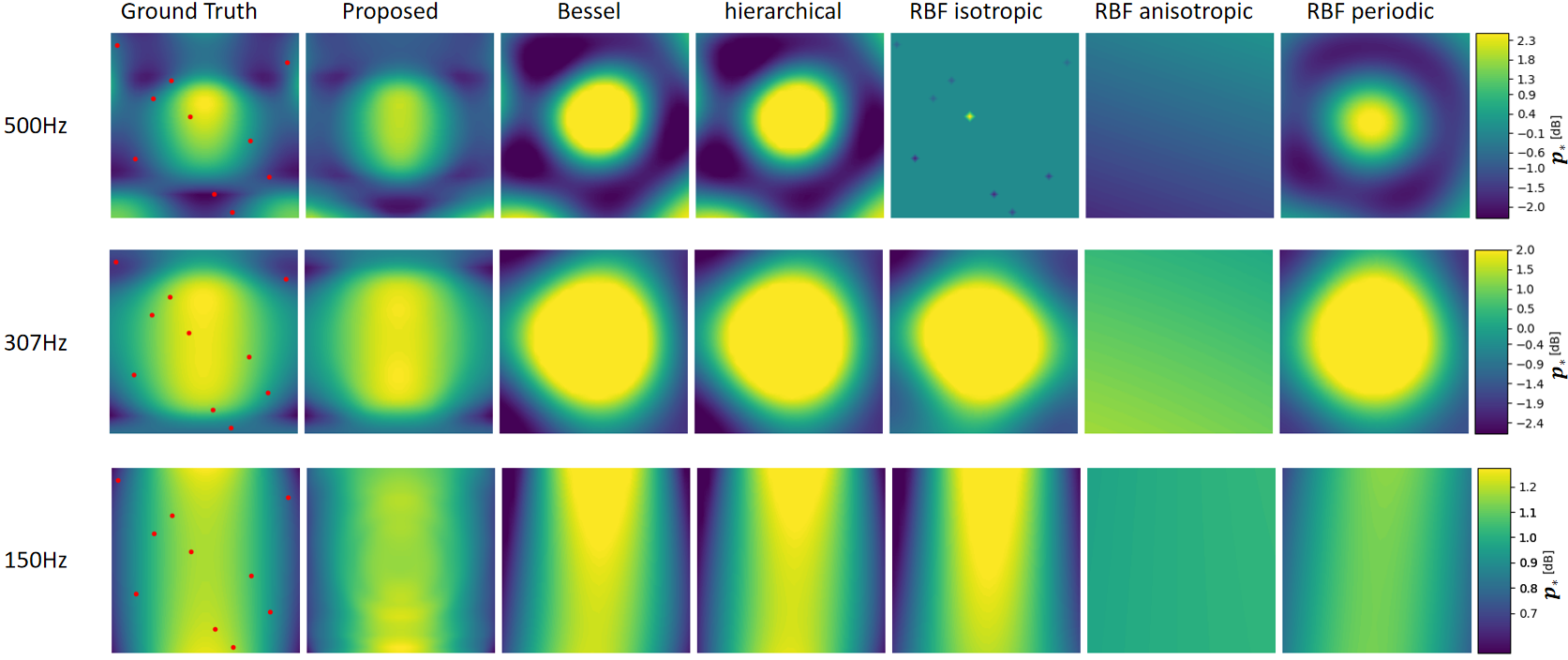}
  \caption{Reconstructed near-field magnitudes of different frequencies given 10 observations arbitrarily placed. The red dots indicate the location used for reconstructing predicted sound field magnitude.
  % As Figure.\ref{fig2} for the near-field of point sources.
  }
 \label{fig3}
\end{figure*}

Table \ref{tab2} shows the mean performance of our proposed model and GPs on a diverse set of 1000 near-fields. To provide a detailed 
reconstruction demonstration, we selected a sound field from the test set for visualization. Figure \ref{fig3} shows the reconstruction of the near-field produced by five point sources evenly distributed at $2\lambda$ m from the center of the reconstructed area. From the figure, we see that the GPs method with existing kernels fails to accurately follow the distance inverse law in terms of the pressure amplitude reconstruction. This discrepancy arises from the mismatch between the kernel functions and the properties of the sound field. Specifically, the magnitude of the reconstructed sound field is relatively small near the sources (i.e., the edge of the reconstruction area), while the magnitude is excessively large at locations further away from the sources (i.e., the center of the reconstruction area). In addition, the kernels are poor for source localization, making it difficult to distinguish the location or even the number of sound sources from Figure \ref{fig3}.

As predicted, the proposed model demonstrates superior performance in accurately reconstructing sources with varying numbers, orientations, and distances, particularly at 500Hz. This outcome highlights the remarkable ability of the proposed model to generalize effectively and reconstruct diverse sound fields.

\begin{table*}[ht]
\caption{The mean of NMSE and MAC of near-field test dataset of different frequencies given 10 observations arbitrarily placed.}\label{tab2}
\begin{tabular*}{\textwidth}{@{\extracolsep\fill}lcccccc}
\toprule%
& \multicolumn{3}{@{}c@{}}{NMSE(dB)} & \multicolumn{3}{@{}c@{}}{MAC} \\\cmidrule{2-4}\cmidrule{5-7}%
Method & 150Hz & 307Hz & 500Hz & 150Hz & 307Hz & 500Hz \\
\midrule
\textbf{Proposed}  & \textbf{-17.8494} & \textbf{-10.5829} & \textbf{-7.0664} & \textbf{0.9872}& \textbf{0.9134} & \textbf{0.8077}\\
Bessel  & -13.7646 & -1.5615  & 2.4769  & 0.9793& 0.9272 & 0.7050\\
hierarchical  &-15.1681 & -1.7293  & 2.2114  & 0.9849 & 0.9294  &0.7083\\
RBF isotropic  & -15.7987 & -1.1729  & 3.1793  &0.9753& 0.0088 & 0.0112\\
RBF anisotropic  &-8.7447 & 3.9365 & 3.6229  &0.9387 & 0.0946  & 0.0019\\
RBF periodic  & -15.1838 & -1.1221 & 3.2672 &0.9866 & 0.9136  & 0.5357\\
\bottomrule
\end{tabular*}
\end{table*}

\subsection*{4.5 RTF Magnitude reconstruction}

RTFs are a crucial component for achieving immersive and interactive sound field reproduction in virtual reality applications~\cite{das2021room}. They represent the frequency-domain representation of RIRs, which typically comprise direct and reverberant components that can be modeled by spherical waves and diffuse fields, respectively ~\cite{fernandez2021reconstruction}. In sections 4.3 and 4.4, we demonstrated the remarkable superiority of our proposed model over the GP method in both near-field and diffuse-field sound field reconstruction. To provide a fair comparison, we further compare our proposed model with a data-driven sound field reconstruction method based on a U-net-like neural network ~\cite{lluis2020sound}. The training process and settings are in line with the original work~\cite{create_dataset}. We employed two simulation methods, the Image-Source Method (ISM) and Modal Theory (MT), to generate RTF datasets. We tested the ability of our model to reconstruct sound fields in simple small-sized rooms, as well as complex rooms with standing waves. Note that the trained networks are not specific to any particular room geometries or wall reflective properties but only leverage the limited set of observations within the reconstruction area of interest, demonstrating the versatility and practicality of our proposed approach.

\subsubsection*{4.5.1 ISM-RTFs dataset}

\begin{figure}[ht!]
  \centering
  \includegraphics[scale=0.5]{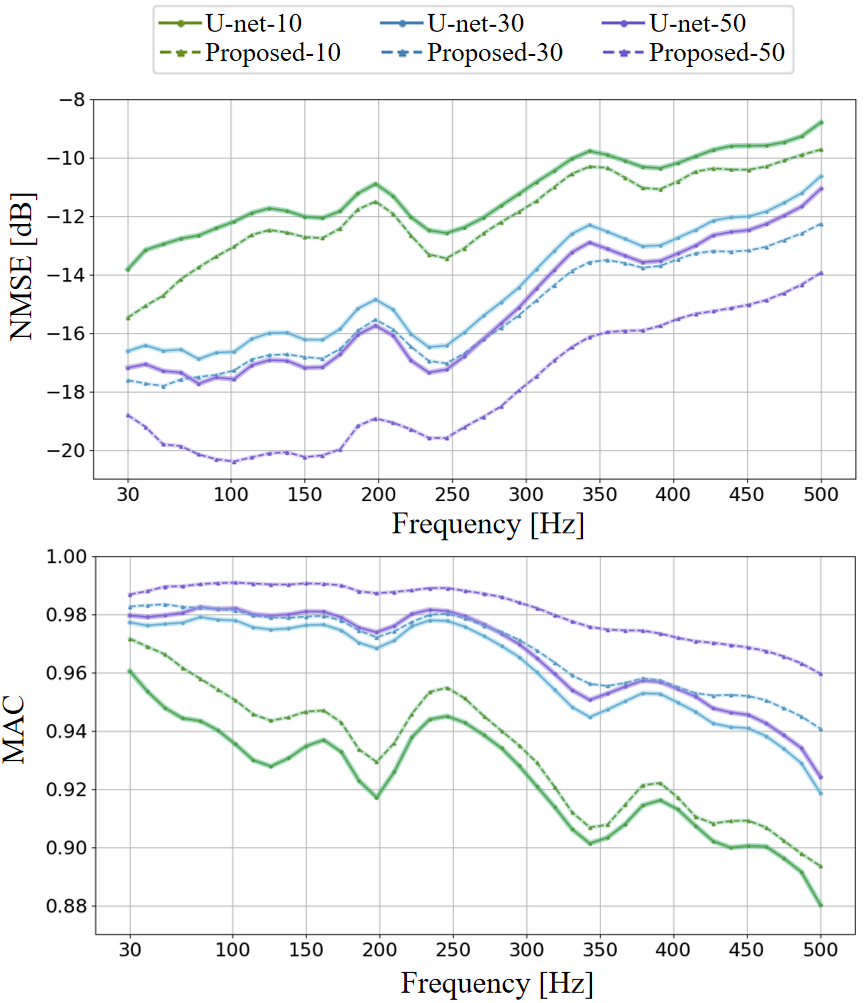}
  \caption{Normalized mean square error (NMSE) in dB and Modal Assurance Criterion (MAC) estimated from ISM's RTF dataset given 10, 30, and 50 observations arbitrarily placed.}
 \label{fig4}
\end{figure}

The ISM for generating RIRs is widely used in sound field reconstruction \cite{fu2022sparse, damiano2021soundfield}, with the RIR generator \cite{habets2006room} being a popular tool due to its simplicity and computational efficiency. The ISM-based approach is well-suited for small room sizes and simple geometries.  In the frequency domain, the generated RTFs are represented as
\begin{equation}\label{29}
p\left(\omega, \mathbf{r} \mid \mathbf{r}_{\mathbf{0}}\right)=\sum_{\beta}^{B} \sum_{\gamma=-\infty}^{\infty} A(\omega) \frac{\mathrm{e}^{j\left(\omega t-k\left\|\mathbf{r}_{\mathbf{\beta}}+\mathbf{r}_{\mathbf{\gamma}}\right\|\right)}}{4 \pi\left\|\mathbf{r}_{\mathbf{\beta}}+\mathbf{r}_{\mathbf{\gamma}}\right\|},
\end{equation}
where $\mathbf{r}_{\mathbf{\beta}}$ are the vectors corresponding to the permutations of $\left(x_{0} \pm x, y_{0} \pm y, z_{0} \pm z\right)$, $\gamma$ is the integer vector triplet $(n_x, n_y, n_z)$ and $\mathbf{r}_{\mathbf{\gamma}}= 2({n_x}{L_x}, {n_y}{}, {n_z}{L_z})$\cite{fernandez2021reconstruction}.

In our simulations, we investigated point source radiation in 2D rooms within the frequency range of [30,500] Hz, where $B=4$ and $z=0$ as specified in Eq.(\ref{29}). We conducted the simulations in 11,000 rectangular rooms with floor areas randomly sampled from 12 m$^2$ to 20 m$^2$. In each room, an omnidirectional source was placed in a uniformly sampled random location. We set reverberation time $T_{60} = 0.4s $, the sampling frequency to $fs = 48$ kHz, and simulate reflections up to the 3rd order. To assess the performance of our proposed model with a limited number of observations, we placed 10, 30, and 50 microphones in a 32 by 32 grid in an arbitrary manner. We used 10,000 and 1,000 rooms for training and testing the model, respectively, from the dataset. We then analyzed the mean performance of the model across these test rooms.

%We simulate point source radiation in the frequency band [30,500] Hz in 2D rooms, where $B=4$ and $z=0$ in the Eq.(\ref{29}). The simulation was conducted in 11000 rectangular rooms, where the floor area is randomly created in the ranges from 12 m$^2$ to 20 m$^2$.  To investigate the performance of the proposed model with very few observations, we arbitrarily placed 10, 30, and 50 microphones in the 32 by 32 grid. For training and testing the model, 10,000 and 1,000 rooms, respectively, were used from the dataset. The mean performance was then analyzed based on these test rooms.

As shown in Figure \ref{fig4}, our proposed algorithm consistently outperforms the U-net model. Specifically, the proposed model achieves similar results with only 30 observations, while U-net requires 50 observations to achieve comparable performance. This improvement can be attributed to the dynamic kernel that incorporates global information more comprehensively than the partial convolution employed in U-net\cite{liu2018image}. This demonstrates the potential of our proposed model to reduce the number of required samples while maintaining its effectiveness.

Furthermore, we observe that the performance of our proposed model improves as the number of available observations increases. Although the performance slightly degrades with increasing frequency, the model still exhibits good performance in reconstructing RTFs in small rooms across most frequencies. These outcomes suggest that our algorithm is effective for reconstructing RTFs in small rooms.

\subsubsection*{4.5.2 MT-RTFs dataset}

In order to investigate the potential of our proposed model for reconstructing complex sound field with standing waves, we generated a dataset using MT~\cite{create_dataset}, i.e., the following equation
\begin{equation}\label{30}
G\left(\mathbf{r}, \mathbf{r}_{0}, w\right) \approx-\frac{1}{V} \sum_{N} \frac{\psi_{N}(\mathbf{r}) \psi_{N}\left(\mathbf{r}_{0}\right)}{(\omega / c)^{2}-\left(\omega_{N} / c\right)^{2}-j \omega / \tau_{N}},
\end{equation}
where $\sum_{N}$ is a triple summation across the modal order in each dimension $(n_x, n_y, n_z)$ of the room, $V$ is the room volume, $\psi_{N}(\cdot)$ is the eigenfunctions  (representing the mode shape), and $\omega_{N}$ denotes eigenfrequencies (representing the resonance frequency). The time constant $\tau_{N}$ represents the characteristic time for a specific mode in a room, and it is a constant obtained by dividing the total sound energy in the room by the sound power absorbed by the walls related to that particular mode. We focus on 2D rectangular rooms in frequency band [30,500] Hz, incorporating all room modes with an eigenfrequency below 600 Hz, and setting $n_z$ to 0 in Eq.(\ref{30}). A reverberation time of $T_{60} = 0.4s $ is assumed. The training and test sets are split, and the room size and sound source location settings are the same as in Section 4.5.1.

\begin{figure}[ht!]
  \centering
  \includegraphics[scale=0.5]{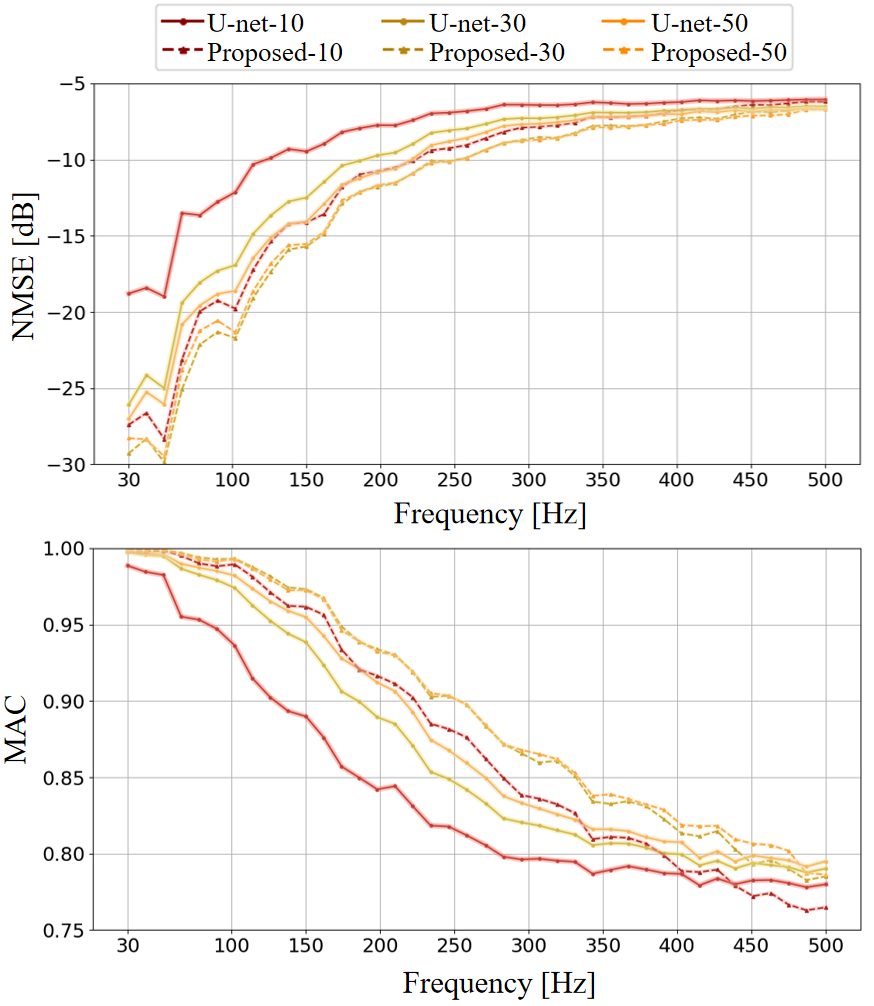}
  \caption{Normalized mean square error (NMSE) in dB and Modal Assurance Criterion (MAC) estimated from MT's RTF dataset given 10, 30, and 50 observations arbitrarily placed.}
 \label{fig5}
\end{figure}

Figure \ref{fig5} depicts the mean performance of the proposed model in reconstructing MT's RTF dataset. The performance of the proposed model given 10, 30, and 50 observations consistently outperforms the U-net, indicating its potential for effectively reconstructing standing waves. Particularly in the low-frequency range, the proposed model exhibits a significant advantage. As the reconstruction frequency approaches the highest eigenfrequency, the complexity of the modes increases, which leads to a decrease in the reconstruction performance. This phenomenon aligns with theoretical expectations, suggesting that a higher number of observations is required to improve robustness and overcome the challenges posed by undersampling~\cite{lluis2020sound,mignot2013room}.

In addition, comparing Figures \ref{fig4} and \ref{fig5}, the method's performance deteriorates with increasing frequency, which is more noticeable in Figures \ref{fig5}. The reason for this phenomenon is the ISM-RTFs dataset is more homogeneous than the MT-RTFs dataset. Specifically, the sound fields generated by IMS are produced in shoebox rooms with image reflections up to the 3rd order.
This indicates a relatively sparse sound field with wavefronts in the space-time domain. Due to the transient nature of the wavefronts, this type of sound field is dense in the frequency domain. In contrast, the sound fields generated by MT are relatively sparse in the modal region of the sound field (up to Schroeder's frequency). As frequency approaches
Schroeder's frequency, the sound fields have increasingly more modes and eventually become diffuse.

\subsection*{4.6 Dynamic kernel visualization}

\begin{figure}[ht!]
  \centering
  \includegraphics[scale=0.42]{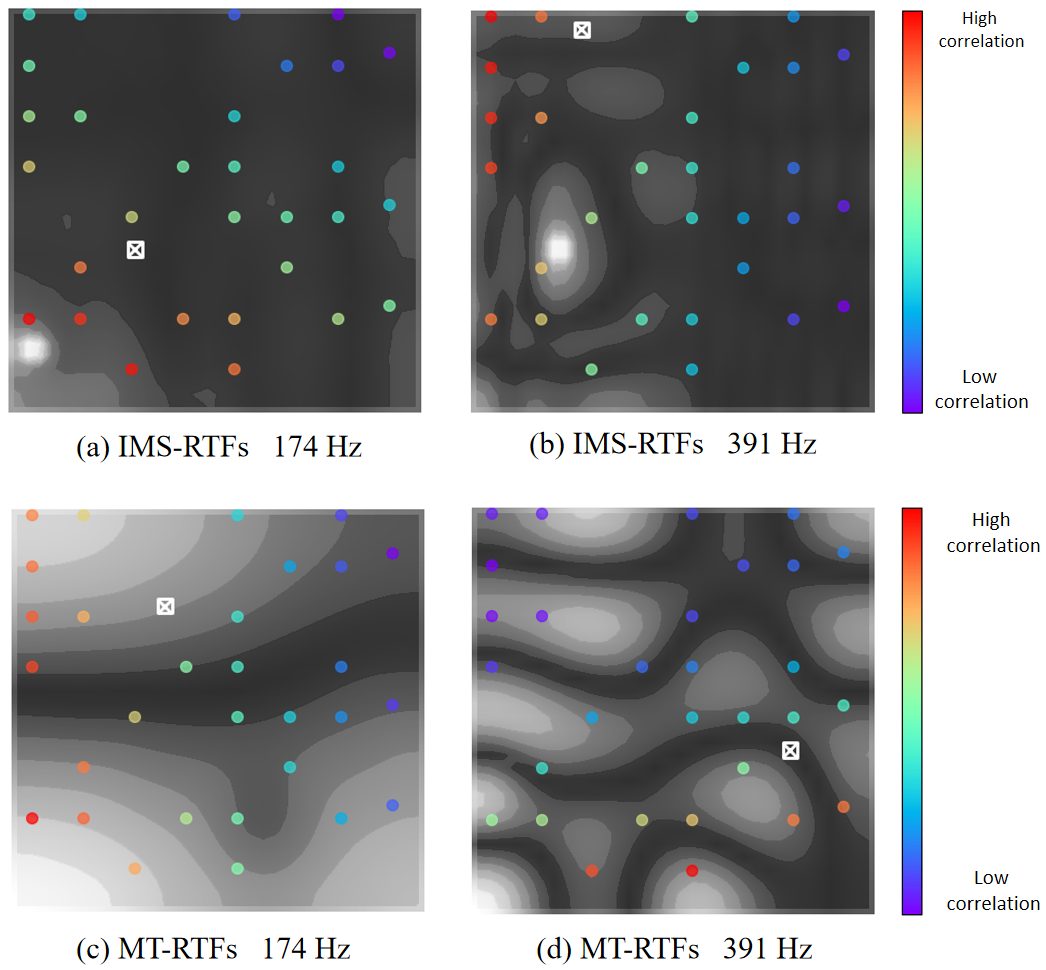}
  \caption{Visualization of spatial correlation of RTFs at a specific frequency. The dots indicate the location of the observations that were used to reconstruct the output of our model, and the white square denotes the target location that needs to predict its magnitude. The color of the dots reflects the strength of the correlation between the observations and the target.}
 \label{fig6}
\end{figure}

In this section, we demonstrate the spatial correlation between observations and target locations using the proposed dynamic kernel Eq.(\ref{19}). We select multiple rooms from both IMS-RTFs and MT-RTFs datasets to visualize the sound field and their spatial correlation at specific frequencies.

Figure \ref{fig6}(a) and (b) demonstrate that for the IMS-RTFs dataset, the correlation is stronger between observations in close proximity to the target location. Additionally, the dynamic kernel assigns relatively more attention to locations where the sound source is situated, i.e., the bottom left of Figure \ref{fig6}(a) and the middle left of Figure \ref{fig6}(b), and less to areas where the sound field characteristics are less prominent, such as the top side of Figure \ref{fig6}(a) and right side of Figure \ref{fig6}(b). This reflects the validity of the dynamic kernel in apportioning attention to the global sound field. Additionally, it provides an explanation for the experimental results in Section 4.3, as the sound field reconstructed by the proposed method reflects the locations of sources.

For the MT-RTFs dataset shown in Figure \ref{fig6}(c) and (d), similar conclusions can be drawn, with closer observations displaying a stronger correlation with the target location. Interestingly, the observations that correlate most strongly with the target point are not in proximity to it but rather at the left bottom of the figure  \ref{fig6}(c) and the bottom of the figure \ref{fig6}(d), where the structural features of the sound field are noticeable. This highlights the dynamic kernel's ability to learn from data. Furthermore, it is apparent that the sound field environment in the MT-RTFs dataset is more intricate than that of the IMS-RTFs dataset at the same frequency. This difference explains the proposed model's performance degradation in reconstructing the MT-RTFs dataset at higher frequencies.

\subsection*{4.7 Model generalization}
%- train with the diverse dataset}
To assess the generalization ability of our model, we combined the four datasets mentioned in Sections 4.3, 4.4, and 4.5 into a diverse dataset for both training and testing. We conducted experiments on four types of sound fields, where 10 observations were arbitrarily placed. In our comparisons between U-net and GPs, we employed the best-performing hierarchical kernel for GPs.

\begin{figure*}[ht!]
  \centering
  \includegraphics[scale=0.3]{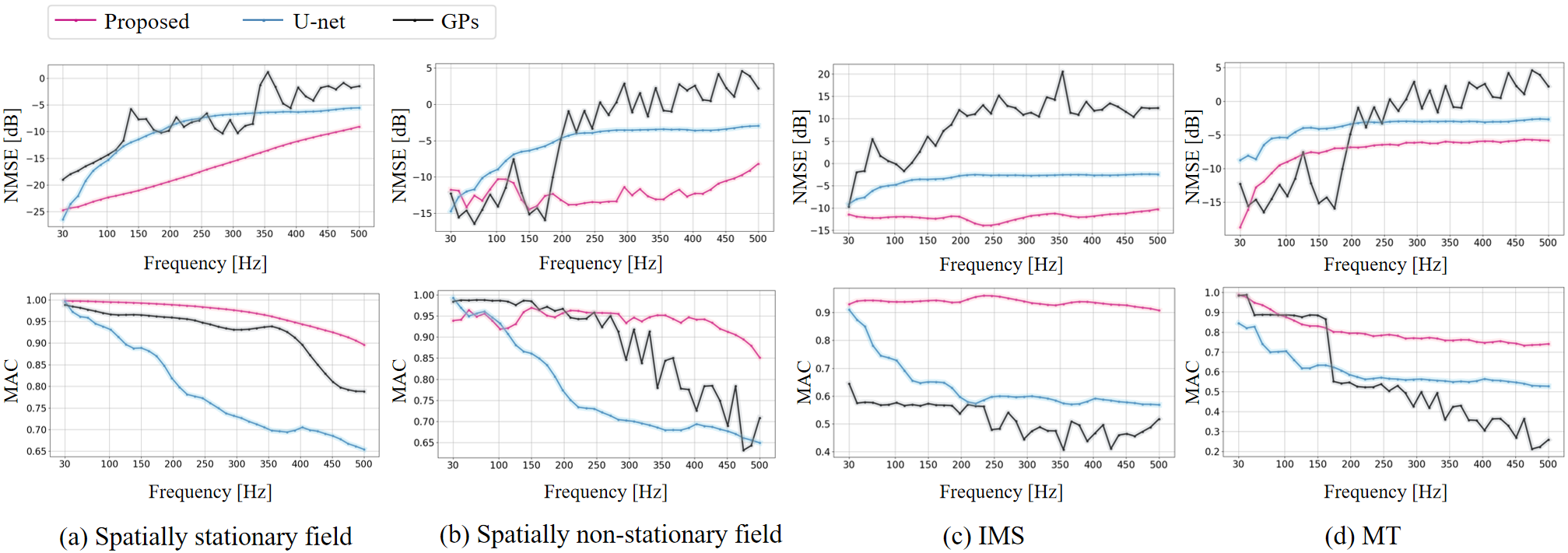}
  \caption{Normalized mean square error (NMSE) in dB and Modal Assurance Criterion (MAC) estimated from four datasets given 10 observations arbitrarily placed.} 
 \label{fig7}
\end{figure*}

As illustrated in Figure \ref{fig7}, we observed a decline in performance for the model trained on the diverse dataset when compared to training on each individual dataset separately. This decline can be attributed to the varying data distributions present in each dataset. However, it is important to note that even with this decline, our proposed model still exhibited strong performance, particularly in terms of robustness at high frequencies.

This outcome serves as a testament to our model’s ability to learn from diverse data and highlights its applicability across various sound field scenarios. While the varying data distributions affected the model’s performance to some extent, our model showcased resilience and delivered notable results, particularly in capturing sound characteristics at higher frequencies.

%In order to verify the generalization ability of the model, we combine the four datasets mentioned in Sections 4.3, 4.4, and 4.5 into a diverse dataset for training and test four types of sound fields given 10 observations arbitrarily placed, separately. We compare U-net and GPs, using the best-performing hierarchical kernel for GPs. As Figure \ref{fig7} shows, due to the varying data distributions in each dataset, the performance of the model trained on the diverse dataset experienced a certain degree of decline compared to training separately on each individual dataset. However, our proposed model still demonstrated strong performance, particularly in terms of robustness at high frequencies. This outcome further underscores the model's ability to learn from the data and its applicability to diverse sound fields.}

\subsection*{4.8 Computational complexity analysis}

Apart from enhancing the accuracy of reconstruction, the proposed model also offers a significant advantage in terms of computational complexity during the inference process. With a model size of 4.3 million parameters, the deterministic inference time is around 0.016 seconds on a Nvidia Tesla K80 GPU. This estimation is based on the observation of 1000 different room predictions. In our experiments, we conducted model training for 300 epochs on the training set. Each type of sound field required approximately 12 hours of training time. The U-net model size is 3.9 million parameters resulting in a deterministic inference time of approximately 0.083 seconds on a Nvidia Tesla K80 GPU. Each type of sound field required approximately 24 hours of training time for 300 epochs.

\section*{5 Conclusion}

In this work, we proposed a novel method that parameterizes GPs using a deep neural network based on Neural Processes. Our method allows for the learning of dynamic kernels from simulated data with the introduction of attention, enabling the method to obtain a kernel that adapts to the acoustic properties of the sound field without many functional design restrictions. Numerical experiment results demonstrate that our proposed method outperforms current methods in terms of reconstructing accuracy for a diverse range of sound fields. Future work involves validating our approach using real-world data and further developing the methodology for complex sound field reconstruction.

%\paragraph*{Sub-sub-sub heading for section}
%\section*{Appendix}
%Text for this section\ldots

%%%%%%%%%%%%%%%%%%%%%%%%%%%%%%%%%%%%%%%%%%%%%%
%%                                          %%
%% Backmatter begins here                   %%
%%                                          %%
%%%%%%%%%%%%%%%%%%%%%%%%%%%%%%%%%%%%%%%%%%%%%%

\begin{backmatter}

\section*{Acknowledgements}

Not applicable.

\section*{Authors' contributions}
WZ and ZL formized and conceptualized the problem. ZL performed the experiments. WZ and TDA supervised the research. All authors read and proved the published version of the manuscript.

\section*{Funding}%% if any
This work was supported in part by the National Natural Science Foundation of China (NSFC) under Grants 61831019 and 62271401.

\section*{Abbreviations}
\textbf{CNN}:Convolutional Neural Networks \\
\textbf{PICNN}:Physics-Informed Convolutional Neural Networks \\
\textbf{ULA}:Uniform Linear Array\\
\textbf{GPs}: Gaussian Processes \\
\textbf{NPs}: Neural Processes \\
\textbf{GELU}: Gaussian Error Linear Unit \\
\textbf{MLP}: Multi-layer perceptron \\
\textbf{CA}: Cross attention\\
\textbf{ELBO}: Evidence lower-bound\\
\textbf{RTF}: Room transfer functions\\
\textbf{NMSE}: Normalized mean square error\\
\textbf{MAC}: Model assurance criteria\\
\textbf{ISM}: Image source method\\
\textbf{MT}: Modal theory

\section*{Availability of data and materials}

The dataset used and analyzed during the current study are not publicly available but are available from the corresponding author on reasonable request.

%\section*{Ethics approval and consent to participate}%% if any Text for this section\ldots
%\section*{Declarations}

\section*{Competing interests}
The authors declare that they have no competing interests.

%\section*{Consent for publication}%% if any
%Text for this section\ldots

%\section*{Authors' information}%% if any
%Text for this section\ldots

%%  After submission of the .TEX file,                     %%
%%  you will be prompted to submit your .BBL file.         %%

% if your bibliography is in bibtex format, use those commands:
\bibliographystyle{bmc-mathphys} % Style BST file (bmc-mathphys, vancouver, spbasic).
\bibliography{bmc_article}      % Bibliography file (usually '*.bib' )
% for author-year bibliography (bmc-mathphys or spbasic)
% a) write to bib file (bmc-mathphys only)
% @settings{label, options="nameyear"}

% or include bibliography directly:
% \begin{thebibliography}
% \bibitem{b1}
% \end{thebibliography}

%%%%%%%%%%%%%%%%%%%%%%%%%%%%%%%%%%%
%%                               %%
%% Figures                       %%
%%                               %%
%% NB: this is for captions and  %%
%% Titles. All graphics must be  %%
%% submitted separately and NOT  %%
%% included in the Tex document  %%
%%                               %%
%%%%%%%%%%%%%%%%%%%%%%%%%%%%%%%%%%%

%%
%% Do not use \listoffigures as most will included as separate files

%%%%%%%%%%%%%%%%%%%%%%%%%%%%%%%%%%%
%%                               %%
%% Tables                        %%
%%                               %%
%%%%%%%%%%%%%%%%%%%%%%%%%%%%%%%%%%%

%%%%%%%%%%%%%%%%%%%%%%%%%%%%%%%%%%%
%%                               %%
%% Additional Files              %%
%%                               %%
%%%%%%%%%%%%%%%%%%%%%%%%%%%%%%%%%%%

%\section*{Additional Files}
% \subsection*{Additional file 1 --- Sample additional file title} Additional file descriptions text (including details of how toview the file, if it is in a non-standard format or the file extension).  This mightrefer to a multi-page table or a figure.\subsection*{Additional file 2 --- Sample additional file title}
  %  Additional file descriptions text.

\end{backmatter}
\end{document}